\documentstyle[12pt,aasms4]{article}
\begin{document}
\def\etal{{\it et al.\/\ }}
\def\today{\ifcase\month
 &  &  &  &  &  \or January &  \or February \or March &  & \or April 
 &  &  &  &  &  \or May &  &  &  \or June &  &  \or July &  &  \or August
 &  &  &  &  &  \or September \or October & \or November \or December\fi
 &  &  &  &  &  \space\number\day, \number\year}
\def\ga{\lower 2pt \hbox{$\, \buildrel {\scriptstyle >}\over{\scriptstyle \sim}\,$}}
\def\la{\lower 2pt \hbox{$\, \buildrel {\scriptstyle <}\over{\scriptstyle \sim}\,$}}

\title{\bf SN 1987A'S CIRCUMSTELLAR ENVELOPE, II: KINEMATICS OF THE THREE
RINGS AND THE DIFFUSE NEBULA$^1$}

\author{Arlin P. S. Crotts$^2$}
\affil{Department of Astronomy, Columbia University, 538 W.~120th St., 
New York, NY~~10027}

\author{Stephen R. Heathcote}
\affil{Cerro Tololo Inter-American Observatory, Casilla 603, La Serena, Chile}

\affil{\bigskip
$^1$ Based in part on observations made with the NASA/ESA {\it Hubble Space
Telescope}, obtained from the data archive at the Space Telescope Science
Institute, which is operated by AURA, Inc., under NASA contract NAS 5-26555.}

\affil{\bigskip
$^2$ Guest observer at the Cerro Tololo Inter-American Observatory, National
Optical Astronomy Observatories, 
which is operated by AURA, Inc., under a cooperative agreement with the NSF}

\authoremail{arlin@astro.columbia.edu}

\begin{abstract}

We present several different measurements of the velocities of structures
within the circumstellar envelope of SN 1987A, including the inner, equatorial
ring (ER), the outer rings (ORs), and the diffuse nebulosity at radii $\la
5$~pc, based on CTIO 4m and $HST$ data.
A comparison of STIS and WFPC2 [N~II]$\lambda$6583 loci for the rings show that
the ER is expanding in radius at (10.5$\pm0.3$)~km~s$^{-1}$ (with the northern
OR expanding along the line of sight at $\sim$26~km~s$^{-1}$, and for the
southern OR, $\sim$23~km~s$^{-1}$).
The best fit to CTIO 4m/echelle spectra of the [N~II]$\lambda$6583 line show
the ORs expanding at $\sim$23~km~s$^{-1}$ along the line of sight.
Accounting for inclination, the best fit to all data for the expansion in
radius of both ORs is 26~km~s$^{-1}$.
The ratio of the ER to the OR velocity is nearly equal to the ratio of the ER 
to the OR radius, so the rings are roughly homologous, all having been created
$\sim$20,000~yr before the SN explosion.
This makes the previously reported, large compositional differences between the
ER and ORs difficult to understand.
Additionally, a grid of longslit 4m/echelle spectra centered on the SN shows
two velocity components over a region roughly coextensive with the outer
circumstellar envelope extending $\sim$5~pc (20 arcsec) from the SN.
One component is blueshifted $\sim$10~km~s$^{-1}$ relative to the systemic
velocity of the SN, while the other is redshifted by a
similar amount.
These features may represent a bipolar flow expanding from the SN, in which
the ORs are propelled 10-15~km~s$^{-1}$ faster than that of the surrounding
envelope into which they propagate.
The kinetic timescale for the entire nebula is $\ga$350,000~yr (and probably
more since material may be accumulating in an outer contact discontinuity).
The kinematics of these different structures constrain possible models for the
evolution of the progenitor and its formation of a mass loss nebula.

\end{abstract}

\clearpage

\section{Introduction}

Supernova 1987A provides a uniquely detailed case for the study of the late
evolution of massive stars with many surprising features.
It is one of the very few SNe in which the progenitor was studied before the
explosion.
The internal composition of the progenitor structure and that of the ejecta
have been the subject of intense scrutiny.
The SN environment is dominated by a mass-loss nebula (or ``circumstellar
envelope'' - CSE) with many different features capable of betraying various
phases and events during the life of the progenitor.
The CSE can in turn be probed through its response to the ionizing flash from
the SN explosion, followed by light echoes in the optical, followed by the
collision with the CSE of the ejecta from the explosion. 

We are studying the spatial structure of this nebula, as well as its velocity
structure seen in narrow emission lines, in order to construct as complete as
possible a history of the mass loss phases of the star.
With both the velocity structure and the full three-dimensional spatial
structure that light echoes reveal, one can interpret observations in terms of
the kinematic ages and therefore the evolutionary phases involved in mass loss.
We are presenting these observations in a series of works detailing the spatial
structure of the inner CSE (Paper I: Crotts, Kunkel \& Heathcote 1995), its
velocity structure (this paper) and the spatial structure of the outer CSE (in
preparation).

While the velocity and distance from the SN of a feature yield a timescale
characteristic of how long ago that material was ejected from the surface of 
the progenitor, these timescales can also be inferred more indirectly from
compositional variations within the CSE: structures arising at different times
result from different layers of the star, and hence should have different
compositions.
Detecting this level of differentiation across the CSE requires the spatial
resolution of $HST$~FOS spectra to compare the rings (Panagia et al.~1996),
indicating much less nitrogen in the outer rings than the inner ring,
suggesting that the outer rings were ejected some $10^4$~yr earlier than the
inner rings whose kinematic age is $\sim 2\times 10^4$~yr (Crotts \& Heathcote
1991).
We shall here test this compositional age measure against the kinematic age of
the outer ring, and attempt to explain possible discrepancies.

\section{Observations}

This is a difficult observational problem best attacked through the combination
of ground-based and $HST$ measurements.
Over the entire central region where they are expected to show the largest
variation in velocity, the ORs are found within only $\sim$0.3 arcsec of the
ER, yet are many times fainter than the ER.
This situation cries out for the high angular resolution of $HST$, but also
requires high spectral resolution (preferably at least $R \approx 30000$) which
is available on $HST$ (GHRS and STIS/MAMA) only in the UV for transitions in
which the ORs are too faint.
Such high spectral resolution is readily available in the optical from the
ground, but seeing limits its usefulness to the portions of the ORs separated
from the ER by more than about 1~arcsec.
As we shall see, the $R\approx 5000$, STIS G750M setup for [N~II]$\lambda$6583
can detect velocity offsets of some ER and OR features if sufficient care is
taken.
Fortunately, for different segments of the ORs, one or the other observational
alternative offers some way of measuring their velocity, and for some these
results can be intercompared.

To look at signals at different spatial scales and spectral resolutions, we
used a total of four different kinds of data: from the ground (CTIO
4m/Echelle), single-slit spectra to study the ER and OR features 1-2~arcsec
from the ER and multi-slit spectra to study fainter features at larger radii,
and on $HST$, STIS spectra of the ER and OR compared to WFPC2 images, the
latter used to fix the angular positions of different features in the STIS
spectrograph slit relative to one another.
To appreciate the relevance of each data set to the problem, it is helpful to
see (Figure 1) the placement of the STIS and CTIO/echelle slits relative to the
rings and adjacent stars.
(``NOR'' and ``SOR'' refer to the northern and southern OR, respectively.)

\subsection{$HST$ STIS G750M Spectra and WFPC2 Images}

The data used here, already described in Sonneborn et al.~(1998) (their Figure
6), were taken on UT 1997 Apr 26, day 3715 after core collapse. 
It was a 643s STIS integration in the G750M mode (6295-6867\AA, resolution $=$
1.02\AA, or 46.5~km~s$^{-1}$ at [N~II]$\lambda6583$).
The 2~arcsec-wide longslit centered on the SN was oriented at PA$=87^\circ$,
nearly parallel to the ER's projected major axis, and hence included the entire
ER as well as much of the ORs, including the segments which might be expected
to be seen expanding most rapidly from the SN along the line of sight.

Because of the wide spectrograph slit used the position of a feature along 
the dispersion axis depends on both its velocity and its position 
perpendicular to the slit.
Thus in order to measure velocities within the rings, we must compare their
dispersed images seen in the STIS spectra to their undispersed locus measured
in WFPC2 images. 
For this purpose we have used an image through the narrow band F658N filter, 
which primarily admits flux in [N~II]$\lambda$6583 obtained on day 3478 
(GO program 6437, Kirshner et al.) which had a total exposure time of 5400s.
The possible motion of the ring emission locus over this 237 day interval
between the STIS and WFPC2 observations should be negligible, less than 0.1
pixel (c.f.~Plait et al.~1995).

\subsection{CTIO 4m Multiple-Order Echelle Spectra}

These data were taken with the same setup used in Crotts \& Heathcote (1991),
using the f/8 Echelle Spectrograph on the CTIO 4m, with the 31.6~l~mm$^{-1}$
echelle grating and KPGL3 cross-disperser (527~l~mm$^{-1}$, blazed at
5500\AA\ in first order), which together deliver a resolving power $R \approx
40000$ (corresponding to 7.5~km~s$^{-1}$) with the 1~arcsec slit used.
We used the red, long-focus camera and red collimator, covering 5530-7050\AA.
We concentrate primarily on the narrow emission line at [N~II]$\lambda6583$.
For the slit decker used, each order is 19 arcsec wide.

Data such as these were taken on numerous occasions between 1989 and 1997, and
we chose the best of these, in terms of good seeing and signal-to-noise ratio
($S/N$) of the OR features.
Those chosen were from UT 1996 Dec 1 (day 3569 after core collapse, shortly
before the $HST$ STIS observation), and have seeing of about 0.8 arcsec FWHM.

\subsection{CTIO 4m Single-Order Longslit Echelle Spectra}

These data are described in Xu \& Crotts (1999).
They employed the 79~l~mm$^{-1}$ echelle grating on the CTIO 4m/Echelle, along
with an order separation filter that selected H$\alpha$ and [N~II] lines,
projected through a set of five slits separated by 12.5~arcsec perpendicular to
their length.
The free velocity range of 120~km~s$^{-1}$ between successive slits was
sufficient to prevent overlap of detectable LMC emission components.
The data considered here were part of a 6$\times$6 arcmin raster of north-south
slits roughly centered on the SN (with one actually passing through the SN),
with additional single, long slits oriented east-west in preliminary exposures
to check that there were no observable components lying beyond the
120~km~s$^{-1}$ no-overlap range.  
These spectra have a velocity resolution of 10~km~s$^{-1}$.
Exposures were obtained for 1~hour in each position, and up to five velocity
components were disentangled from each spectrum, analyzed at each point on a
grid with 13~arcsec sampling north-south and 12.5~arcsec east-west.
These data were taken several years before the others, on days 2160-2162.

\section{Results}

These four datasets are combined to map the velocity field of SN~1987A's CSE.
The ground-based data provides good measurements of the velocities of the ORs
north and south of the SN, which are missed by STIS, but because of limited
spatial resolution are less useful for the portions of the ORs closer to the
ER, which STIS covers well.
The emission from the portion of the sky containing the outer CSE, however, has
too low an emission measure for $HST$ to study easily, but is seen more easily
by a telescope with a faster f-ratio.
We consider each of these measurements in turn.

\subsection{Comparison of OR Loci from WFPC2 and STIS}

The G750M STIS spectra of Sonneborn et al.~(1998), their Figure 6, show
excellent images of the ER and a portion of the NOR and SOR in
[N~II]$\lambda$$\lambda$6548, 6583 and H$\alpha$ (as well as several other
transitions) taken on day 3715, which can be compared to nearly contemporaneous
images from WFPC2.
The [N~II]$\lambda$6583 line has by far the greatest $S/N$ (with the H$\alpha$
signal being smeared by thermal broadening.)~
The difference between the positions of features in the dispersed STIS data 
compared to the direct WFPC2 image allows us to measure the relative velocities
between different parts of the CSE.
Unfortunately, the region of the NOR seen within the ER is largely obscured by
the SN ejecta and in part by the blue-shifted smear of the ``hot spot''
(Sonneborn et al.~1998, Michael et al.~1998) on the ER overlapping the NOR.
Nonetheless, the entire ER and a large portion of the northern SOR and southern
NOR can be compared between the STIS and WFPC2 data in this way.

Figure 2 shows the results of the best bilinear fit of the G750M STIS locus of
the ER to the position of ER from the day 3478 WFPC2 exposure with the F658N
filter.
The loci in each image are defined as the set of points of maximum brightness
defined along radial crosscuts through the SN.
(Note that the scale in one dimension is expanded to make the ER appear
roughly circular.)~
This shows a close correspondence between the two loci with the STIS points
following the WFPC2 track, matching most features in great detail, to within an
r.m.s.~difference of 0.12 STIS pixels.
The anamorphic magnification of this STIS mode at this wavelength (Bowers et
al.~1998) is 0.92786 (compressed in the dispersion direction).
In comparison, the STIS-to-WFPC2 fit requires a 0.91058 compression along the
dispersion, corresponding to a difference in velocity of $0.56\pm 0.02$ STIS
pixels, or $0.31\pm 0.01$\AA\ between the northern and southern side of the ER,
in the sense that the southern side is redshifted.
This implies a velocity shear of 14.3~km~s$^{-1}$ across the slit, which
corresponds (since the slit is well-aligned with the ER) to a radial expansion
velocity i.e.~outward from the SN, of $10.5\pm 0.3$~km~s$^{-1}$.
This is a new, independent measurement of the ER expansion velocity, but it is
in good agreement with some previous values (10.3~km~s$^{-1}$: Crotts \&
Heathcote 1991, Cumming 1994) and is not far from others (11~km~s$^{-1}$:
Panagia et al.~1996, 8~km~s$^{-1}$: Meaburn et al.~1995).

Figure 3a shows the position of the STIS and WFPC2 OR signals after alignment
of the ER signals as in Figure 2.
The residual offset of the SOR and especially the NOR is a clear sign of the
velocity difference between the ORs and the ER centroid.
Intercalibrating the data in this way eliminates any possible sources of
systematic error due to uncertainties in the standard STIS and WFPC2
platescales, at the cost of only slightly increasing the random error due to
the uncertainty of aligning the two ER signals.
The useful sample of the ORs excludes those portions where they approach the
ER, and also where they run nearly parallel to the dispersion axis.
Additionally, the SOR signal is corrupted over a large extent by the continuum
from Star 3.
The uncontaminated samples are shown in Figure 3a, and with the SOR signal
magnified in Figure 3b. 
This remaining sample permits one useful measurement along the SOR at a point
1.36~arcsec east of the SN (labeled ``a'' in Figure 1), and three measurements
of comparable accuracy along the NOR, 1.04, 1.40 and 1.68~arcsec west (points
``b'' through ``d'' in Figure 1).
These yield velocity values of 21.5~km~s$^{-1}$ (blueshifted) and 22.3, 24.6
and 21.1~km~s$^{-1}$ (redshifted), respectively, relative to the
289~km~s$^{-1}$ heliocentric redshift of SN~1987A, with $1 \sigma$ errors of
about 2~km~s$^{-1}$, except for the last point, with an error of about
3~km~s$^{-1}$.

\subsection{Ring Velocities from [N II]$\lambda$6583 Echelle Spectra}

As discussed by Crotts \& Heathcote (1991), the velocity field of the ER, best
traced by [N II]$\lambda$6583, has a centroid at 289~km~s$^{-1}$, with little
velocity shear across the ER at PA$=60^\circ$, almost parallel to the ER major
axis, but with strong velocity shears on either side, at PA$=20^\circ$ and
$130^\circ$.
There are additional, fainter emission patches at larger radius or velocity
offsets from the centroid.
Occasionally such faint sources correspond to transient spots (Crotts, Kunkel
\& McCarthy 1989, Hanuschik 1990, Crotts \& Heathcote 1991, Cumming \& Meikle
1993, Cumming 1994), but at the time of these observations no such spots were
revealed in optical imaging.
These fainter, outrigger sources observed here are almost certainly due to the
ORs.

Figure 4 shows the portion of the echelle order centered on the [N~II]$\lambda
6583$ line, showing the broad emission lines from the SN and stellar continua
extending horizontally, and the LMC interstellar [N~II] emission extending
vertically.
The bright patch in the center is emission from the CSE.
The underlying ISM is fit linearly in the spatial dimension, interpolated
across the CSE, and subtracted off.
The spectral continuum and broad emission is fit quadratically, and likewise
interpolated and subtracted.

This subtraction produces the residual distribution of flux shown in the
contour plot of the echellogram in Figure 5a (with a vertical scale of
0.26~arcsec and horizontal scale of 3.50~km~s$^{-1}$ per pixel, at
[N~II]$\lambda$6583).
There are notable protrusions from the nearly elliptical flux contours,
particularly to the south at roughly the centroid velocity and to the north 
slightly to the red of centroid.
Furthermore, the flux contours at low levels become more rectangular, as if
flux might be appearing at the corners.

\subsubsection{Inner Ring Fit}

We know the flux distribution in [N~II]$\lambda$6583 of both the ER and ORs
from the HST images, and have good constraints on the global velocity structure
of the ER.
Furthermore, comparing the width of the ER feature in the [N~II] and H$\alpha$
echellograms, we see a significant difference most likely due to thermal
broadening.
Accounting for the different atomic masses producing the lines we find the
temperature of the gas at day 3569 to be roughly $T=5000$K (to within roughly
20\%).
(This compares to other temperature measurements at late times: $T \approx
17000$K on day 1278, from [O II] and [N II] recombination [Wang 1991, Sonneborn
et al.~1997]; 55000K on $\sim$day 300 from [O III] [Wampler \& Richichi 1989].
The lower temperature we measure on day 3569 may be due to the cooling of the
gas, or due to H $\alpha$ and [N~II] emission arising from different gas.
Recombination model calculations predict $T \approx 5000$K throughout the ER by
day 2000 [Lundqvist \& Fransson 1996].)~
We can also measure the spatial and spectral line spread functions of the
spectrograph from the width of the stellar continuum and night sky lines in the
echellogram.
Given these parameters we can model the image of the CSE produced by the
spectrograph on the detector and construct the synthetic echellogram shown in
Figure 5b.
At high flux levels, it appears very similar to the actual data, but at low
flux levels differs by the absence of the extensions mentioned above.

The width of the spatial and spectral line spread functions, temperature, and 
placement of the slit are allowed to vary over a grid of values spanning
approximately $\pm$10\% in each of these parameters, in order to produce a
family of model echellograms, and the minimum $\chi^2$ (in terms of the
square-root of counts per pixel), within the region covered by non-zero flux
from the model.
The resulting best fit model is shown in Figure 5b, and the difference between
the data and fit shown in Figure 5c.
This corresponds to a best fit of 9.5~km~s$^{-1}$ in radial expansion velocity
of the ring away from the SN, with a probable error of about 5-10\% (which is
difficult to estimate given systematic errors in other parameters).
Note that the number of contours in the residual features is several times
fewer than in the original, Figure 5a, and that the contour levels are five
times finer in Figure 5c, so that the extreme residual features are only about
7\% of the original peak flux, and cover regions on the scale of single pixels,
so that residual features within the region of the model are at most about 1\%
of the total.
% total 106400 ADU = 138300 e.  Sqrt = 371 e = 0.27% of total.
This compares to expected Poisson errors from photon counting statistics at the 
0.3\% level, so a much better fit would be difficult to achieve.

\subsubsection{Outer Ring Signals}

Outside the region modeled in \S 3.2.1, several significant residuals appear:
to the north and south at roughly the centroid velocity, and to higher and
lower velocities at roughly the centroid position along the slit.
%centroid at (16.1, 15.8),  dx = 3.496 km/s,  dy = 0.26 arcsec
%-v: 2994 ADU at (09.9, 14.3):  2.8% at -21.7 km/s, 0.39 arcsec N (N SOR)
% S: 2931 ADU at (16.9, 24.2):  2.8% at +02.8 km/s, 2.18 arcsec S (S SOR)
% N: 2079 ADU at (17.3, 08.2):  2.0% at +04.2 km/s, 1.98 arcsec N (N NOR)
%+v: 1159 ADU at (21.9, 16.0):  1.1% at +20.3 km/s, 0.05 arcsec S (S NOR)
These four residuals are centered at
($-$21.7~km~s$^{-1}$, 0.39 arcsec north),
($+ $2.8~km~s$^{-1}$, 2.18 arcsec south),
($+ $4.2~km~s$^{-1}$, 1.98 arcsec north), and
($+$20.3~km~s$^{-1}$, 0.05 arcsec south),
and have flux amounting to 2.8\%, 2.8\%, 2.0\% and 1.1\%, of the total,
respectively.
Note that these radii along PA$=20^\circ$ correspond closely to those measured
from the WFPC2 image for the northern SOR, southern SOR, northern NOR, and
southern NOR respectively:
%u3gh0401t_c0frot20:xcenter@397.3;NSOR@415.1,SSOR@345.3,NNOR@446.3,SNOR@395.3
0.71 arcsec north, 2.08 arcsec south, 1.96 arcsec north, and 0.08 arcsec south,
with the first and fourth measurements being somewhat uncertain due to 
confusion with the ER and SN, respectively.
These positions agree between the echelle and $HST$ datasets to within about
0.2 arcsec r.m.s., less than an echelle pixel (0.26 arcsec).

We can use the relative strength of the ER and ORs in the WFPC2 images to
predict relatively how much light is entering the echelle spectrograph slit.
From the STIS G750M data, we know that the OR flux in the vicinity of the ER
(but beyond $\sim$0.3~arcsec to avoid being overwhelmed by the ER) are
2.5-3.5 times stronger in the [N~II]$\lambda$6583 line than in H$\alpha$.
Additionally, the throughput at the SN redshift in the WFPC2 F658N filter is
3.8 times greater at [N~II]$\lambda$6583 than at H$\alpha$, hence the OR signal
in the WFPC2 F658N image is 92\% due to [N~II].
This implies that even if the H$\alpha$ flux were to vary by 100\% of its
strength relative to the [N~II]$\lambda$6583 flux at different points around
the ORs, the flux of the OR [N~II]$\lambda$6583 flux can still be predicted to
within 9\%.

%center @ (397.3, 371.6), bg = 1.0 ADU/pix
%ER:   1722*(60.86-1) - 224*(26.56-1) = 97000.
%SSOR: 492*(4.85-1) = 1900 (= 2.0%).
%NNOR: 555*(4.197-1) - 25*(2.143-1) = 1750 (= 1.8%).
The flux values inferred from WFPC2 for the south SOR is 2.0\% and for the
north NOR is 1.8\% of the ER signal within the slit, close to the values seen
in the echellogram.
The brightness of the north SOR and south NOR is difficult to determine in the
WFPC2 image, again because of the contributions by the ER and SN.
Given that most of the signal in the residual echelle features corresponds to
the actual OR fluxes, it would appear that errors in the flux could only
decenter the signal by at most about 1 pixel in velocity, or about
4~km~s$^{-1}$.
We will take this as the probable error on the echelle OR measurements.

\subsection{Kinematic Model of Ring Velocities}

To interpret these OR velocity measurements, we compute a simple model for OR
loci in a homologous velocity flow, without imposing any constraint that the OR
and ER velocities should compare, and then require that the projected shape of
each OR be consistent with its observed shape and location.
For the purpose of computing homologous velocities, we make the approximation
that all three rings are coaxial.
For the NOR, which appears nearly elliptical, we adopt an inclination angle
set by assuming that the ring is really circular and is only seen as an ellipse
in projection.
This inclination angle is 42$^\circ$.7 (Burrows et al.~1995) with an error that
we infer to be about 1$^\circ$.
In contrast, the ER has an inclination measured variously as 43$^\circ \pm 3$
(Jakobsen et al.~1991), 45$^\circ$.3 (Burrows et al.~1995), or 44$^\circ.0 \pm
1.0$ (Plait et al.~1995).
The SOR is not truly elliptical, but the best fit ellipse corresponds to an
inclination of 31$^\circ$ (Burrows et al.~1995).
For the SOR we adopt two alternative geometries: A) that it is inclined by
31$^\circ$, and B) that it is inclined by 45$^\circ$, parallel to the ER, but
distorted from a circle.
Perhaps neither alternative is correct, but these at least span a reasonable
range of possibilities.

Under these assumptions, the NOR is expanding outwards from the SN at
26.1~km~s$^{-1}$, while the SOR expands at either A) 25.5~km~s$^{-1}$, or B)
26.3~km~s$^{-1}$ (along the minor axis), according to a best fit to the
observed velocities from STIS+WFPC2 and the CTIO 4m/Echelle.
These fits are shown in Figures 6a, b and c (for the NOR and SOR models A and
B, respectively).
All three fits are reasonable, with reduced $\chi^2$ near unity.
SOR model B (31$^\circ$ inclination) fits slightly better than model A.
The 1$\sigma$ errors on all of these derived expansion velocities are about
2~km~s$^{-1}$.
(Note that the above results appear consistent with those quoted by Cumming \&
Lundqvist [1994].)

These velocities should be compared the radial distance of the ORs from the
SN in order to derive a kinematic timescale.
For the sake of discussion, we adopt a semimajor axis length of 1.7~arcsec for
both ORs.
(We measure a slightly smaller value than the semimajor axis lengths found by
Burrows et al.~[1995]: 1.77 and 1.84 arcsec.)~
For a distance to SN 1987A of 50 kpc, these models correspond to OR distances
from the SN of 0.58~pc for the NOR, 0.52~pc for the SOR (model A), or 0.56~pc
for the SOR (model B, along the minor axis).
Compared to the corresponding expansion velocities, these correspond to
kinematic timescales of 21700, 19900 and 20800~yr, respectively, indicating
the time in which the ORs would have coasted at their current velocities to
their current positions.
Similarly, the ER, with a semimajor axis of 0.86~arcsec (Plait et al.~1995) or
0.21~pc, would require 19500 years expanding at 10.5~km~s$^{-1}$.
Clearly, these kinematic ages are consistent within the errors (of $\approx$
11\%).

\subsection{Spatial Extent of Longslit Velocity Components}

The longslit survey by Xu \& Crotts (1999) of the 6 arcmin square around
SN~1987A was intended to observe structure on scales considerably larger than
the spatial sampling of 13 arcsec, but nonetheless provides some information on
the CSE of the SN.
As seen in light echoes (Crotts \& Kunkel 1991, Crotts 1999), there is an echo
feature apparently centered on the SN extending some 5~pc towards Earth,
corresponding to an angular scale of 20 arcsec.
A search for features on this scale in the longslit data set reveals two
features (Figure 7).
One, centered near 279~km~s$^{-1}$ (10~km~s$^{-1}$ blueshift with respect to
the SN) is found between 25~arcsec east and 13~arcsec west, and between
18~arcsec north and 4~arcsec south of the SN.
A second feature at 301~km~s$^{-1}$ (12~km~s$^{-1}$ redshift) extends between
13 and 38~arcsec west, and 4~arcsec north and 15~arcsec south of the SN.
These appear to form a possibly bipolar, or double-lobed, nebula, with the
near side to the northeast, assuming material is being ejected.
The axis of symmetry of this structure would lie along $PA \approx 50^\circ$.
These features are not perfectly coherent in velocity, however.
While the $S/N$ detection level of individual segments of these features is
low, there appear to be significant gradients in velocity, so that the
differences from the SN velocity centroid can extend up to 15~km~s$^{-1}$.

It is evident from Figure 7, however, that patches of roughly this spatial
scale and velocity coherence exist in locations far away from the SN, such that
there is a significant probability that one or both features are chance
superpositions in projection against the SN.
One possible method of investigating the association of these two features with
the SN is to determine if their elemental abundances are more similar to the
anomalous values found in the rings (e.g.~Lundqvist \& Fransson 1996, Sonneborn
et al. 1998), or those typical in the LMC.
We are currently pursuing this test of the origin of this gas.

\section{Discussion}

Why do the kinematic and compositional timescales (Panagia et al.~1996)
disagree?
Of course, the simple kinematic timescale $t_{kin}=r/v$ does not account for
the acceleration of the RSG wind by the lower density (and not directly
observed) BSG wind, so that at a given radius the winds' interface i.e.~the
freshly shocked RSG material, is actually expanding faster than the RSG wind
just exterior to the shock.
Since this exterior wind has presumably coasted freely since leaving the
progenitor, $t_{kin}$ for this material is close to the actual time $t_{em}$
since the material left the progenitor.
However, this may not be so for the material actually observed.
In comparing $t_{kin}$ for different parts of the CSE we wish to know
whether the ratio of $t_{kin}/t_{em}$ is constant throughout.
Recent models of interacting winds within a wind-compressed disk outflow
(Collins et al.~1999) can lead to velocity fields over the winds' interface
which produce a nearly constant $t_{kin}/t_{em}$ throughout (Collins et
al.~1999, Fig.~2 and A.~Frank, private communication), while simultaneously
producing a shape for the CSE consistent with that observed.
Other models (Blondin \& Lundqvist 1993, Martin \& Arnett 1995) have a more
complicated velocity flow, in which $t_{kin}/t_{em}$ is nonuniform even on
scales as small as $10^{17}$~cm.
Thus while a constant $t_{kin}/t_{em}$ is consistent with the known morphology
of the CSE, it is by no means assured.

In contrast, the compositional ratios computed by Panagia et al.~(1996) are
highly time-dependent, since the system is far from equilibrium while still
recovering from the pulse of EUV photons emitted during the first hours of the
SN explosion.
While the elemental ratios can be recovered from a time-dependent analysis of
line ratios, the particular line ratios arising from the same gas can vary by
large amounts over the course of time (Lundqvist and Fransson 1991).
The Panagia et al.~study does not sufficiently explore these time-dependent
effects.
It is unclear that a compositional difference between the ER and ORs can be
disentangled from these effects given the single epoch of data and relatively
simple models considered, as a unique solution to the elemental abundances
requires knowledge of the temperature and ionization shortly after it was
heated and ionized by the SN flash (P.~Lundqvist, private communication).

If the extended two-lobed nebula apparently in Figure 7 is indeed associated
with SN 1987A, its expansion velocity is 10-15~km~s$^{-1}$ slower than the OR
velocities, but roughly equal to the ER expansion velocity.
If this nebula is associated with the SN, what is its three-dimensional
configuration?
This is difficult to know certainly given the absence of light echoes from the
outer regions of this double-lobed object.
This absence might simply be due to insufficient light travel time to cover the
whole structure, which would probably require 30~yr or more.
In two-dimensional projection, the correspondence of this structure to gas at
smaller radii is not obvious.
Its symmetry axis ($PA \approx 50^\circ$) is far from the axis of the ring
system ($PA \approx 10^\circ$) or its equatorial plane.
Since this nebula does extend significantly to the north and south of the SN,
however, if it does surround the SN, it likely envelops the ORs, too.
In this case, the ORs are expanding at $\sim$25~km~s$^{-1}$ into a medium
expanding at only 10-15~km~s$^{-1}$, and are therefore being accelerated by the
BSG wind into the slower RSG wind medium.
It is not clear that the ER has been accelerated by a proportionally large
fraction of its own velocity, implying that it might be younger than the ORs.
This runs in the same sense as the inequality between ER and OR ages inferred
by Panagia et al.

Regardless of the nature of the extended nebulosity, it is evident that the
inner portions of the CSE are expanding at different velocities.
This is different from at least the initial conditions assumed in all
simulations of the SN's CSE, with the exception of Collins et al.
Other simulations tend to produce an asymmetric boundary between the BSG and
RSG winds with an isotropy in matter density, not velocity.
It is interesting that both can be probed using not only narrow-line emission,
but also light echoes, which trace dust density.
Furthermore, if the CSE terminates in a contact discontinuity against an
interstellar bubble of constant pressure (Chevalier \& Emmering 1989), the
shape of this contact discontinuity is a measure of the momentum flux as a
function of angle around the SN, since pressure must be matched along the
discontinuity's surface.
Thus in directions from the SN of higher outward ram pressure $\rho v^2$ at a
given radius, the wind will be forced to a halt at a larger radius.
This will be investigated in Paper III.

Regardless of the shape of the CSE, its size scale combined with the outflow
velocity, especially where gas is simply coasting and not being accelerated
along a wind interface, provides an estimate of the minimum age of the CSE.
It is only a lower limit because mass has accumulated in the contact
discontinuity.
A CSE radius of 5~pc and a velocity scale of $\sim$15~km~s$^{-1}$ implies a
minimum CSE lifetime of $\sim$350000~yr.
This compares to model predictions to the duration of the RSG phase of SN 1987A
of about 190000~yr (Woosley, Pinto \& Ensman 1988) minus a terminal BSG phase
beginning 10000-20000~yr before explosion (Woosley et al.~1988, Saio, Kato \&
Nomoto 1988).
While the terminal BSG phase lifetime is in agreement with observations, the
RSG phase appears too short in the models.
This would imply that the SN progenitor depleted its core helium supply earlier
than expected.

Further work is needed to explain the production of these three rings.
Lloyd, O'Brien and Kahn (1995) interpret the rings in terms of interacting
winds from the progenitor in the presence of a companion star.
Also, Chevalier and Dwarkadas (1995) propose that the rings arose from the
collision of the BSG and RSG winds, with the further action of ionization by
the progenitor, but no complete interacting-wind model has yet to be
formulated.
Models (Meyer 1997, Soker 1999) that naturally explain the formation of a
three-ring system fail in other important respects, either in not predicting
(Meyer 1997) the cavity interior to the three rings, as evidenced by the rapid
fall and only gradual rise of the radio signal from ejecta/CSE interaction
(Turtle et al.~1987, Staveley-Smith 1993), or in failing to incorporate (Soker
1999) the diffuse nebulosity seen in the outer CSE (Crotts \& Kunkel 1991,
Crotts 1998) and between the rings (Crotts et al.~1995, see also Lundqvist \&
Fransson 1996).
The eventual model, in order to be successful, must now account for the motions
seen in the rings and beyond.

\acknowledgments

The authors would like to thank Peter Lundqvist and the anonymous referee for
helpful comments.
This research was supported by grant AST 90-22586 from the NSF, STScI grant
AR-05789.01-94A and NASA LTSA grant NAG5-3502 to A.P.S.C.

\newpage
\noindent
FIGURES:

\noindent
Figure 1: A schematic layout of the three rings, traced from the day 3478 WFPC2
F658N image, showing the inner, equatorial ring (ER), southern outer ring (SOR)
and northern outer ring (NOR).
Also shown are the positions of Stars 2 and 3, and the SN (central dot).
Overlaying these are the locations of the three CTIO 4m/echelle spectrograph
slits (1.0~arcsec wide at PA 20, 60 and 130), and the edges of the day 3715
STIS G750M slit (2.0~arcsec wide).
The labels ``a'' through ``d'' rest immediately above the STIS samples referred
to in the text and used to measure OR velocities.
The scale in terms of arcseconds and light-years (for $D_{SN} = 50$~kpc) are
shown.

\noindent
Figure 2: The loci of the day 3715 STIS G750M data of the ER (squares) fit to
overlay the day 3478 WFPC2 F658N image (solid curve).
This bilinear fit is compared to the anamorphic distortion of the STIS G750M
setup to yield the velocity shear across the ER, which corresponds to radial
expansion away from the SN of $10.5\pm 0.3$~km~s$^{-1}$.
Note that the scales on the two axes are unequal, to account for the ring's
inclination.

\noindent
Figure 3a shows the position of the STIS and WFPC2 OR signals in relation to
the ER loci once the ER signals are aligned as in Figure 2.
The offset of the SOR and especially the NOR even at the dispersion position
corresponding to the ER centroid (dashed line) is a clear sign of the velocity
offset of the ORs versus the ER centroid.
The NOR can be split into three roughly equal segments, 1.04, 1.40 and 
1.68~arcsec west of the ER centroid, with velocity offsets between the two
data sets corresponding to 22.3, 24.6 and 21.1~km~s$^{-1}$ (redshifted).

\noindent
Figure 3b shows the same data as in Figure 3a, but with the SOR signal
magnified.
A single measurement of SOR velocity is possible, at a point 1.36~arcsec east
of the SN, yielding 21.5~km~s$^{-1}$ (blueshifted).

\noindent
Figure 4 shows the portion of the echelle order centered on the [N~II]$\lambda
6583$ line, showing the broad emission lines from the SN and stellar continua
extending horizontally, and the LMC interstellar emission in [N~II] extending
vertically.
The bright patch in the center is emission from the CSE.
This data is for PA$=20^\circ$, with north to the top.
Wavelength (vacuum, heliocentric) is shown along the abscissa (modified
slightly by structure in the CSE perpendicular to the slit).
These data have 8~km~s$^{-1}$ velocity resolution.

\noindent
Figure 5a shows the same data as in Figure 4, once the underlying ISM has been
fit linearly in the spatial dimension, interpolated across the CSE, and
subtracted off.
The spectral continuum and broad emission have been fit quadratically, and
likewise interpolated and subtracted.
Echellogram pixels are denoted by the small tickmarks on both sides of both
axes.
The vertical pixel scale is 0.26~arcsec and the horizontal scale is
3.50~km~s$^{-1}$ at [N II]$\lambda$6583).
There are notable protrusions from the nearly elliptical flux contours,
particularly to the south at roughly the centroid velocity and to the north 
slightly to the red of centroid, and flux contours at low levels become more
rectangular, as if flux might be appearing at the corners.

\noindent
Figure 5b shows a fit to brighter flux levels by adjusting for the temperature
of the gas, velocity gradient, seeing, spectral resolution, and placement of
the slit.
It reproduces most features seen in the actual data (Fig.~5b) except for the
extensions at large separations and velocity differences.

\noindent
Figure 5c shows the resulting residual between the actual data and the best
fit.
Note that the number of contours in the residual features is several times
fewer than in the original, Figure 5a, and that the contour levels are five
times finer in Figure 5c, so that the extreme residual features are only about
7\% of the original peak flux, and cover regions on the scale of single pixels,
so that residual features within the region of the model are at most about 1\%
of the total.
This compares to count Poisson errors expected at the 0.3\% level, so is close
to the smallest possible residual.

\noindent
Figure 6a shows the best fit to the measured NOR velocities from both
STIS+WFPC2 and CTIO 4m/Echelle data, assuming that the ring is intrinsically
round.
In Figures 6a-c, the velocity flow is assumed to be homologous.
Under these assumptions, the NOR is expanding radially outwards from the SN
at 26.1~km~s$^{-1}$.

\noindent
Figure 6b shows the best fit to the measured SOR velocities from both
STIS+WFPC2 and CTIO 4m/Echelle data, assuming that the ring is inclined by
$31^\circ$, the angle required to force the semimajor and deprojected semiminor
axis to the same lengths.
Under these assumptions, the SOR is expanding radially at 25.5~km~s$^{-1}$.

\noindent
Figure 6c shows the best fit to the measured SOR velocities from both
STIS+WFPC2 and CTIO 4m/Echelle data, assuming that the ring is inclined by
same value as the ER, at $45^\circ$, but distorted from a circle.
Under these assumptions, the SOR is expanding radially at 26.3~km~s$^{-1}$
(along the observed minor axis).

\noindent
Figure 7 shows a 70 arcsec by 40 arcsec array of longslit echelle spectrograms
from a field centered on SN 1987A (the bright spot in the center), in the
[N II]$\lambda$6583 line.
The data were collected five slits at a time as part of a larger (6~arcmin
$\times$ 6~arcmin) survey, and four such exposures are shown here, separated by
bold black lines.
The free velocity range between successive slits is 120~km~s$^{-1}$, with
10~km~s$^{-1}$ velocity resolution, and the slits are separated by 12.5~arcsec
on the sky.
North is up, east is to the left (from slit to slit), and higher velocities
(between slits) is to the right, as indicated by the larger set of arrows.
Two features, centered $-$10~km~s$^{-1}$ and $+$12~km~s$^{-1}$ with respect to
the SN, are found between two slits east and one slit west, and between one to
three slits west of the SN, respectively.
The first feature is found between the vertical pairs of white arrows, and the
second between pairs of black arrows.
Note that other features of similar size are also seen in different locations
in the field, casting some doubt on the association of the two features with
the SN.


\begin{references}

\noindent Bowers, C.~\& Baum, S.~1998, STIS Instrument Science Report 98-23:
``Plate Scales, Anamorphic Distortion \& Dispersion: CCD Modes''

\noindent Blondin, J.M.~\& Lundqvist, P.~1993, ApJ, 405, 337

\noindent Burrows, C.J., et al.~1995, ApJ, 452, 680

\noindent Chevalier, R.A.~\& Dwarkadas, V.V.~1995, ApJ, 452, L45

\noindent Chevalier, R.A.~\& Emmering, R.T.~1989, ApJ, 342, L75

\noindent Collins, T.J.B., Frank, A., Bjorkman, J.E.~\& Livio, M.~1999, ApJ,
512, 322

\noindent Crotts, A.P.S.~1999, in ESO/CTIO/LCO Symp.~{\it SN 1987A: Ten Years
After}, eds.~N.~Suntzeff \& M.~Phillips, in press

\noindent Crotts, A.P.S.~\& Heathcote, S.R.~1991, Nature, 350, 683

\noindent Crotts, A.P.S.~\& Kunkel, W.E.~1991, ApJ, 366, L73

\noindent Crotts, A.P.S., Kunkel, W.E.~\& McCarthy, P.J.~1989, ApJ, 347, L61

\noindent Cumming, R.J.~1994, Ph.D thesis, Imperial College, London

\noindent Cumming, R.J.~\& Lundqvist, P.~1994, in {\it Advances in Stellar
Evolution}, eds.~R.T.~Rood \& A.~Renzini (Cambridge U.~Press, Cambridge), 297

\noindent Cumming, R.J.~\& Meikle, W.P.S.~1993, MNRAS, 262, 689

\noindent Hanuschik, R.W.~1990, A\&A, 237, 12

\noindent Jakobsen, P., et al.~1991, ApJ, 369, L63

\noindent Lloyd, H.M, O'Brien, T.J.~\& Kahn, F.D.~1995, MNRAS, 273, L19

\noindent Lundqvist, P.~\& Fransson, C.~1991, ApJ, 380, 575

\noindent Lundqvist, P.~\& Fransson, C.~1996, ApJ, 464, 924

\noindent Martin, C.L.~\& Arnett, D.~1995, ApJ, 447, 378

\noindent Meaburn, J., Bryce, M.~\& Holloway, A.J.~1995, Ap\&SS, 233, 75

\noindent Meyer, F.~1997, MNRAS, 285, L11

\noindent Michael, E., McCray, R., Borkowski, K.J., Pun, C.S.J.~\& Sonneborn,
G.~1998, ApJ, 492, L143

\noindent Panagia, N., Scuderi, S., Gilmozzi, R., Challis, P.M., Garnavich,
P.M.~\& Kirshner, R.P.~1996, ApJ, 459, L17

\noindent Plait, P.C., Lundqvist, P., Chevalier, R.A.~\& Kirshner, R.P.~1995,
ApJ, 439, 730

\noindent Saio, H., Kato, M.~\& Nomoto, K.~1988, ApJ, 331, 388

\noindent Soker, N.~1999, MNRAS, 303, 611

\noindent Sonneborn, G., et al.~1997, ApJ, 477, 848

\noindent Sonneborn, G., et al.~1998, ApJ, 492, L139

\noindent Staveley-Smith, L.~et al.~1993, Nature, 366, 136

\noindent Turtle, A.J, Campbell-Wilson, D., Bunton, J.D., Jauncey, D.L.~\&
Kesteven, M.J.~1987, Nature, 327, 38

\noindent Wampler, E.J.~\& Richichi, A.~1989, A\&A, 217, 31

\noindent Woosley, S.E., Pinto, P.A.~\& Ensman, L.~1988, ApJ, 324, 466

\noindent Xu, J.~\& Crotts, A.P.S.~1999, ApJ, 511, 262

\noindent 

\end{references}
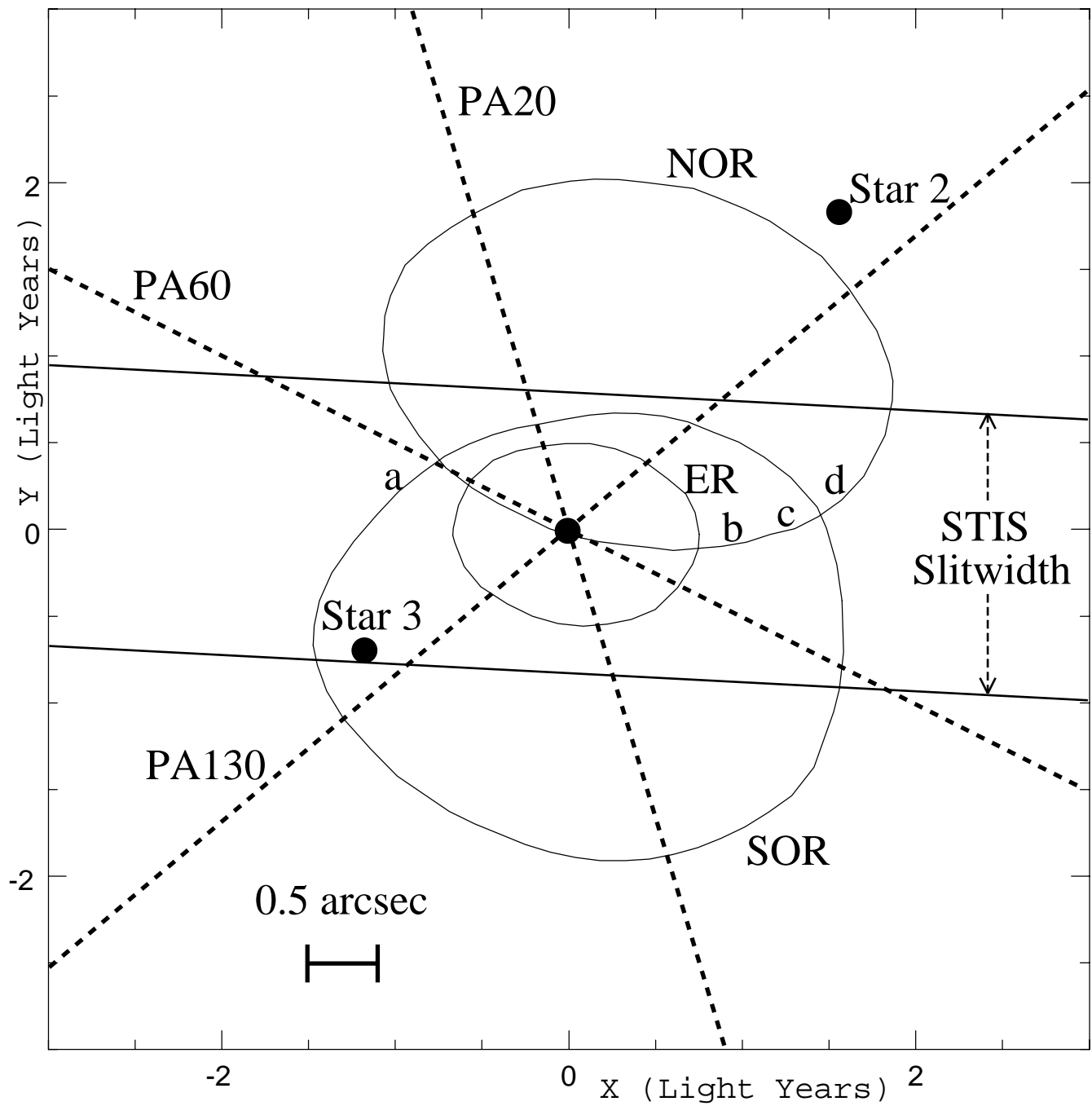
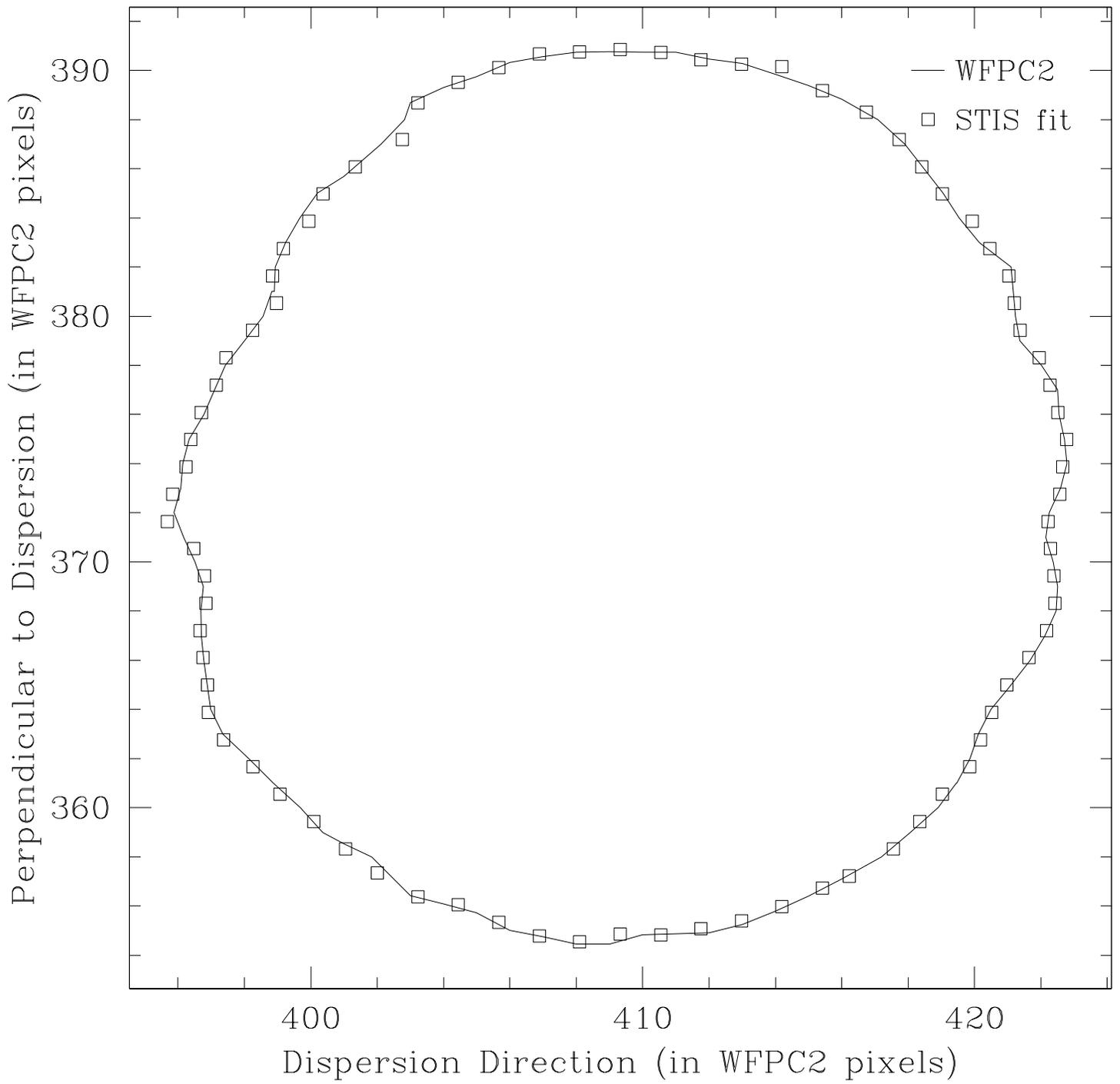
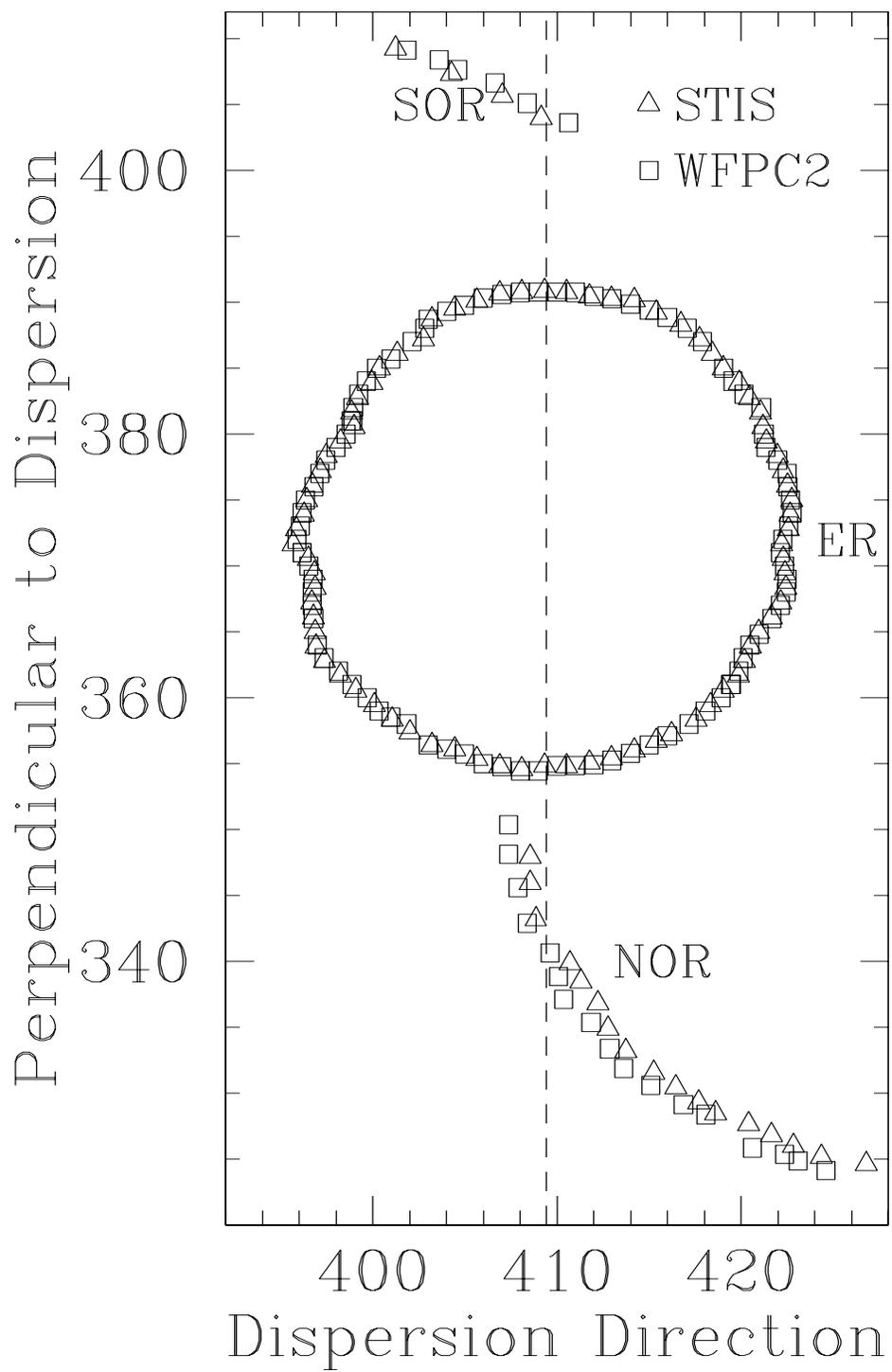
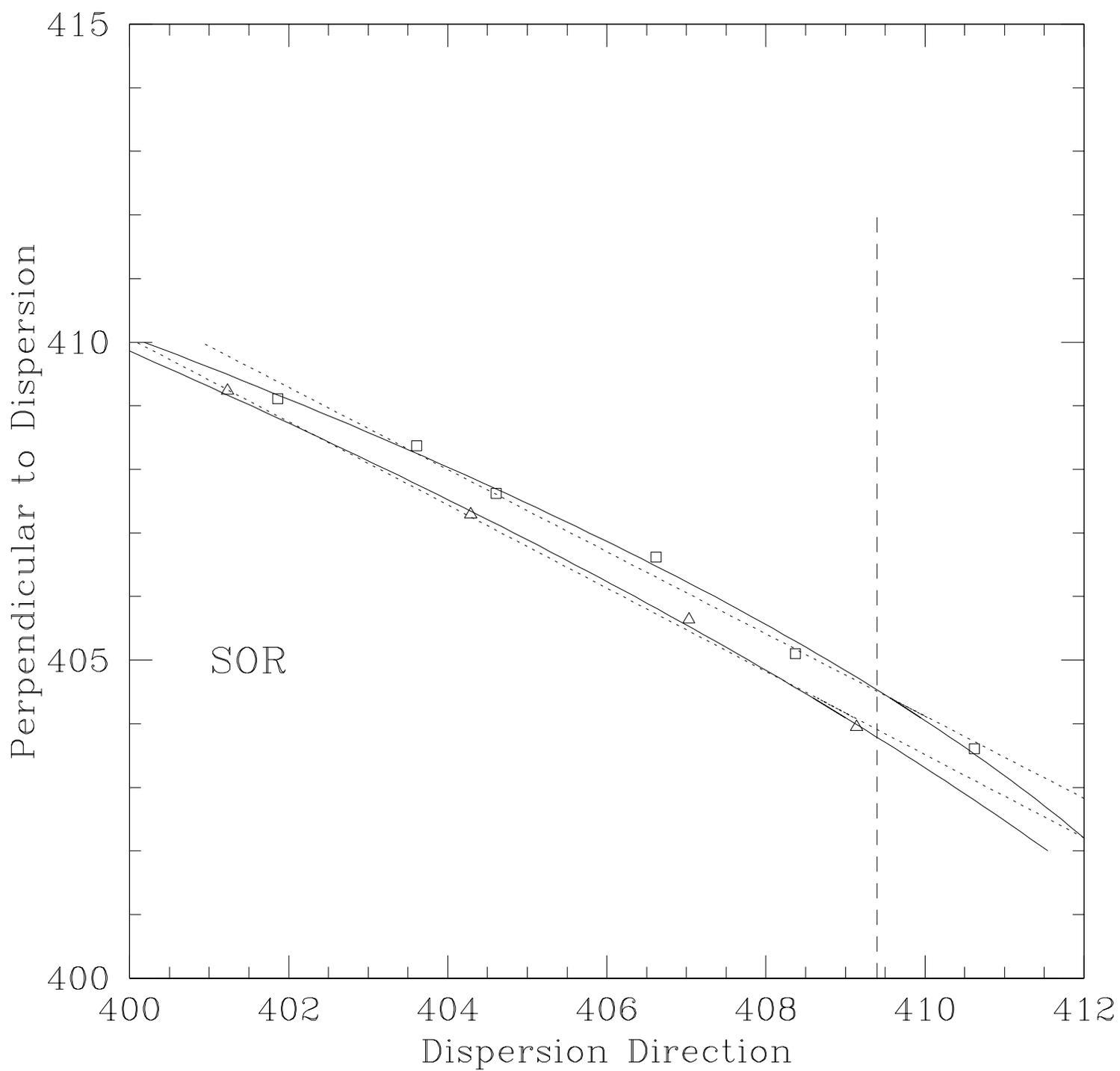
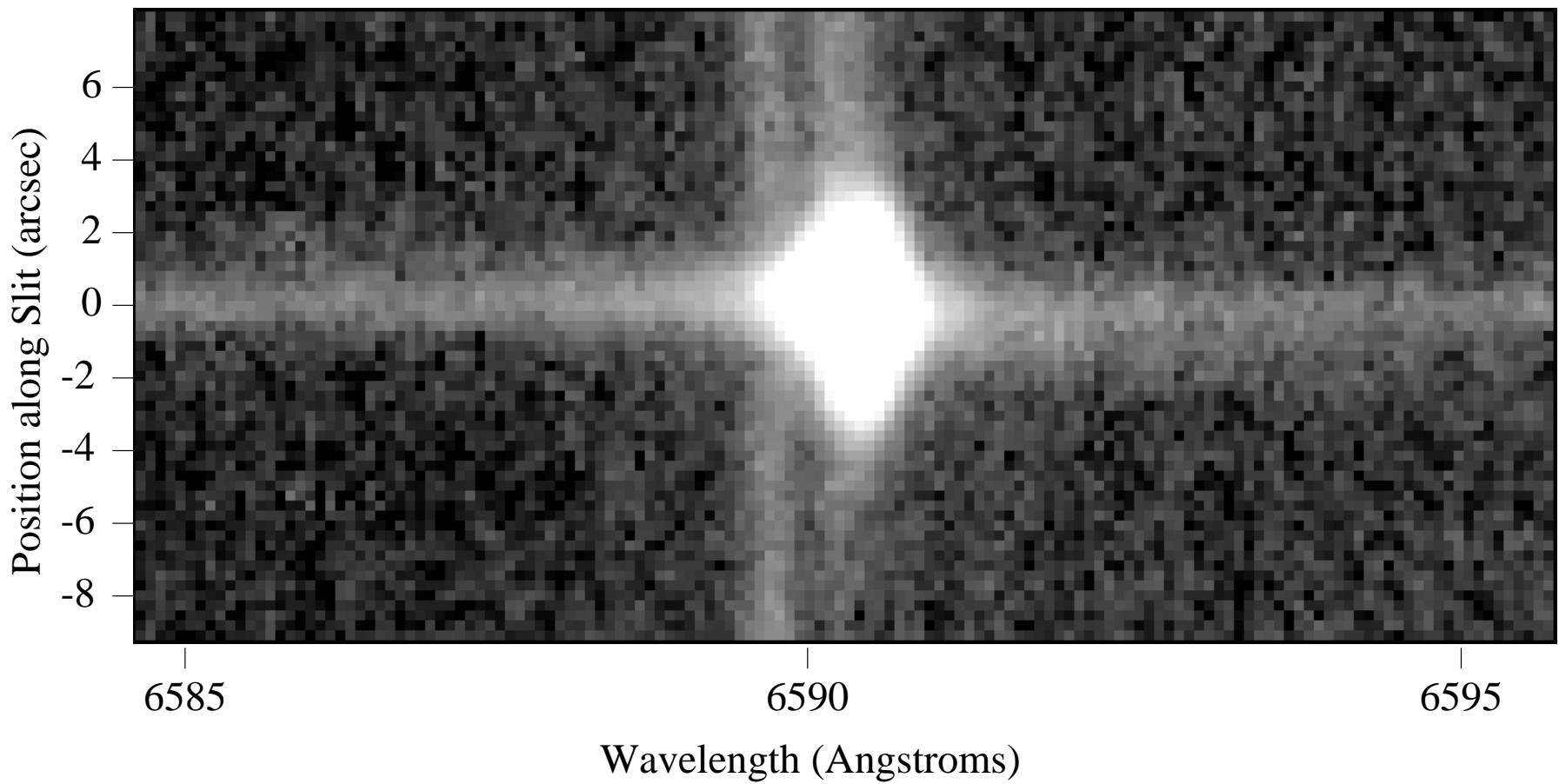
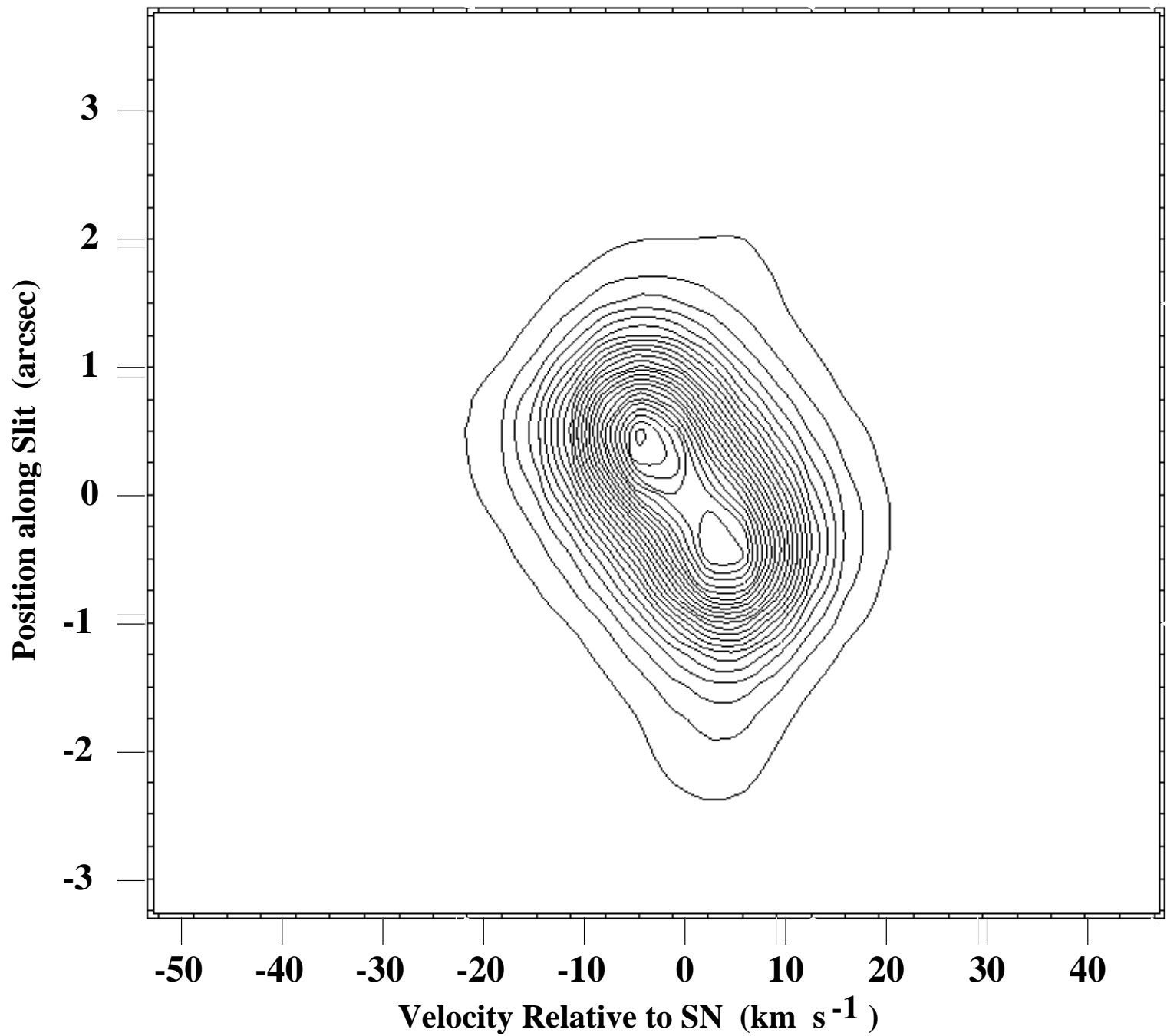
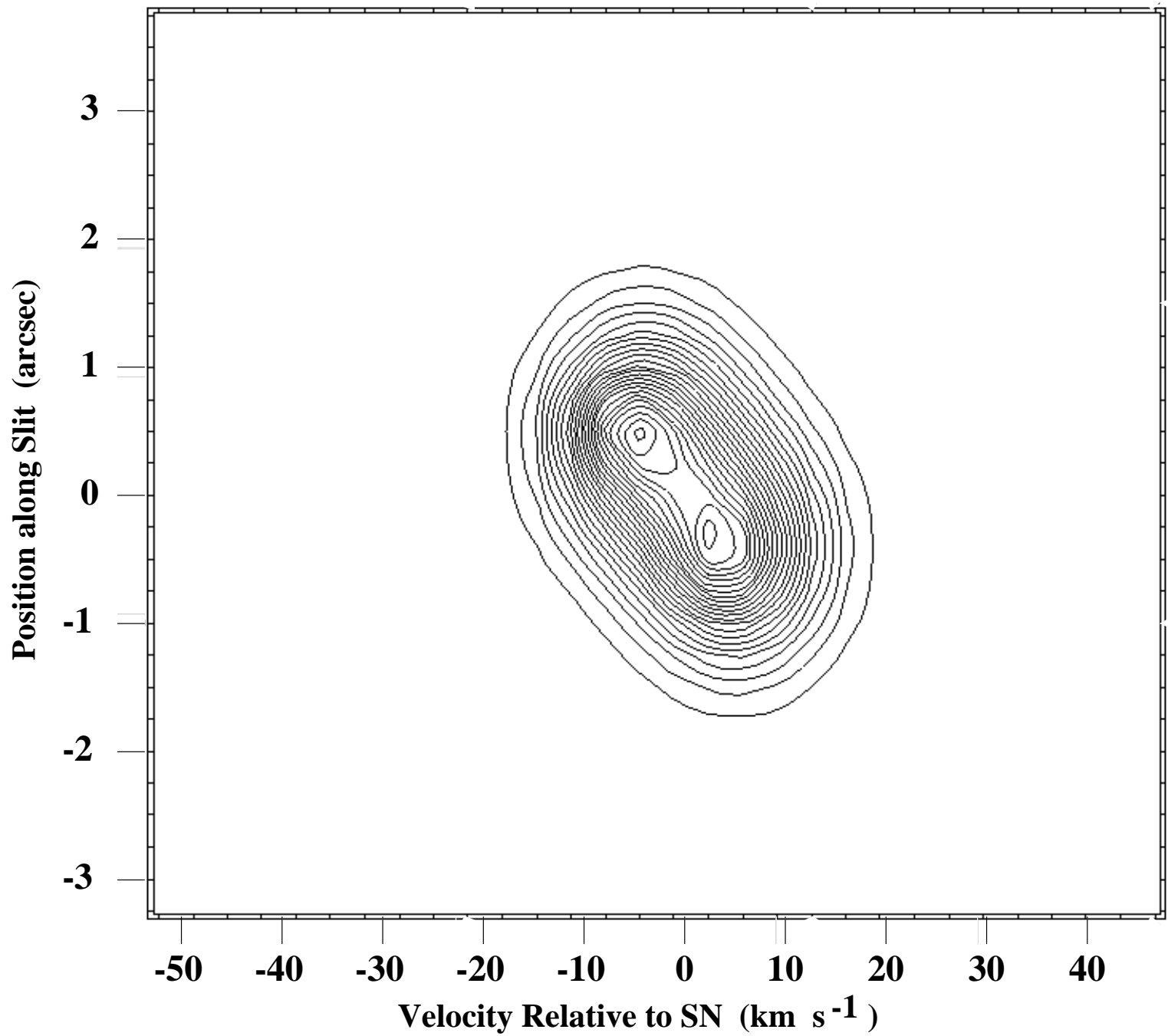
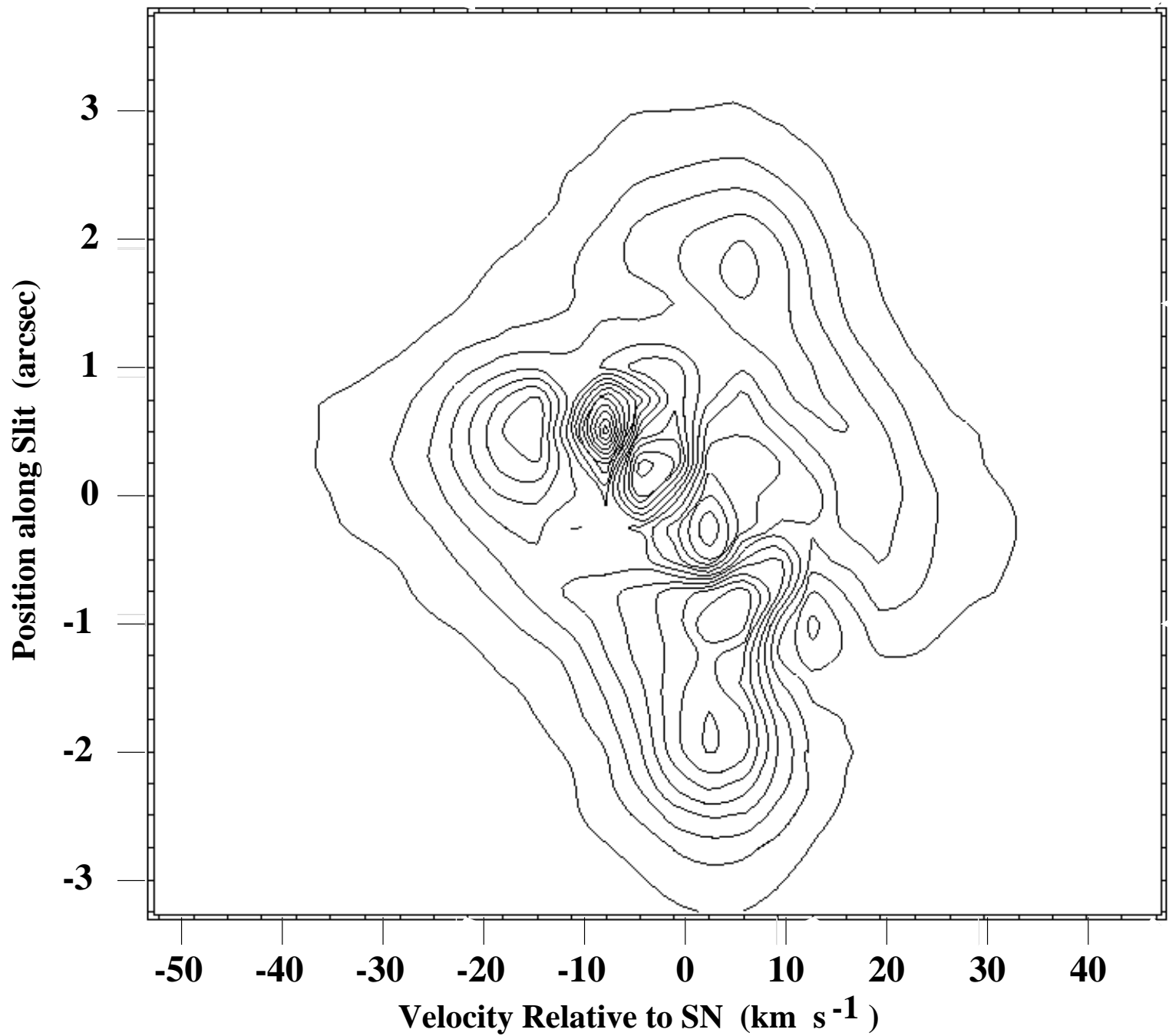
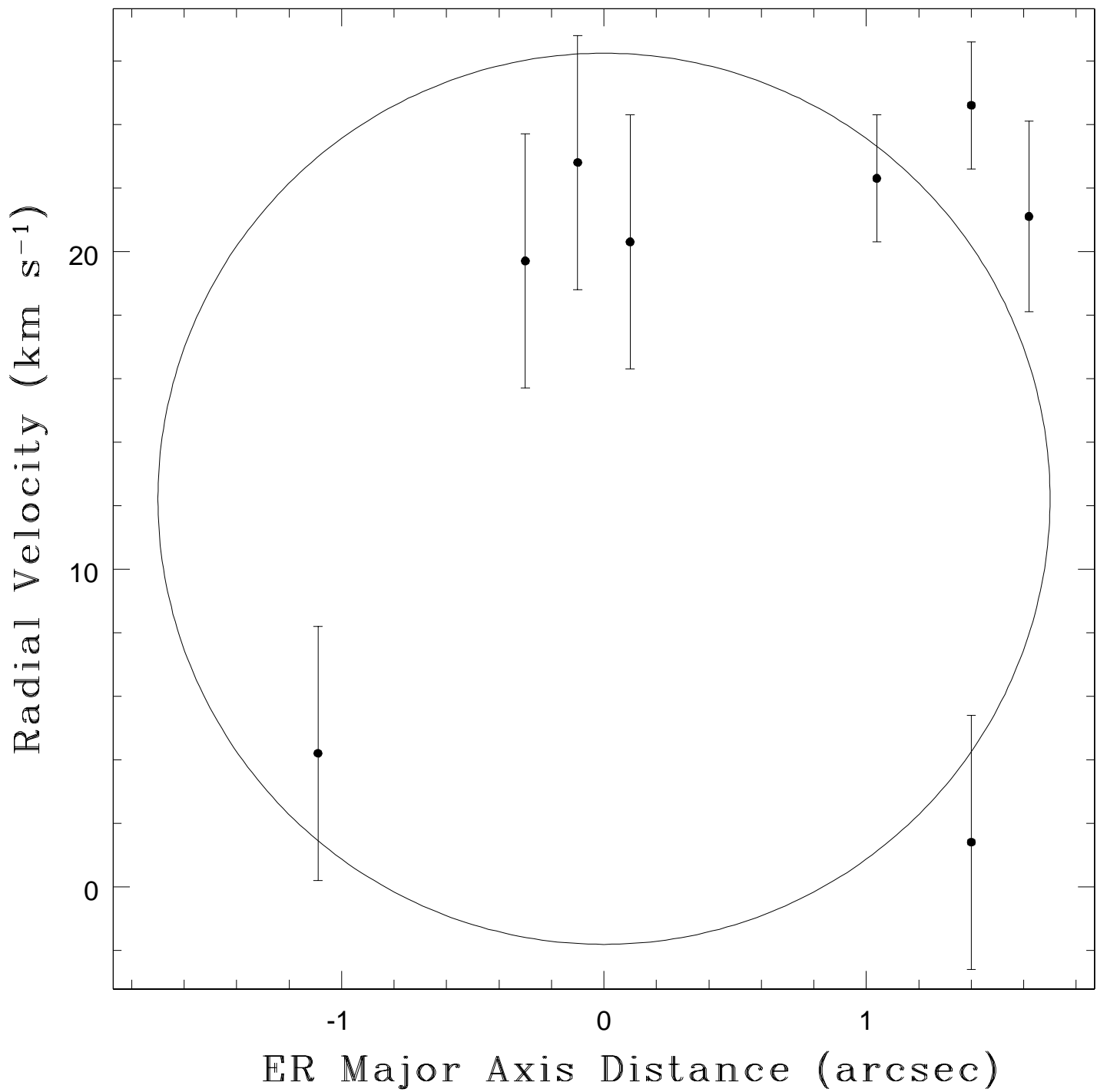
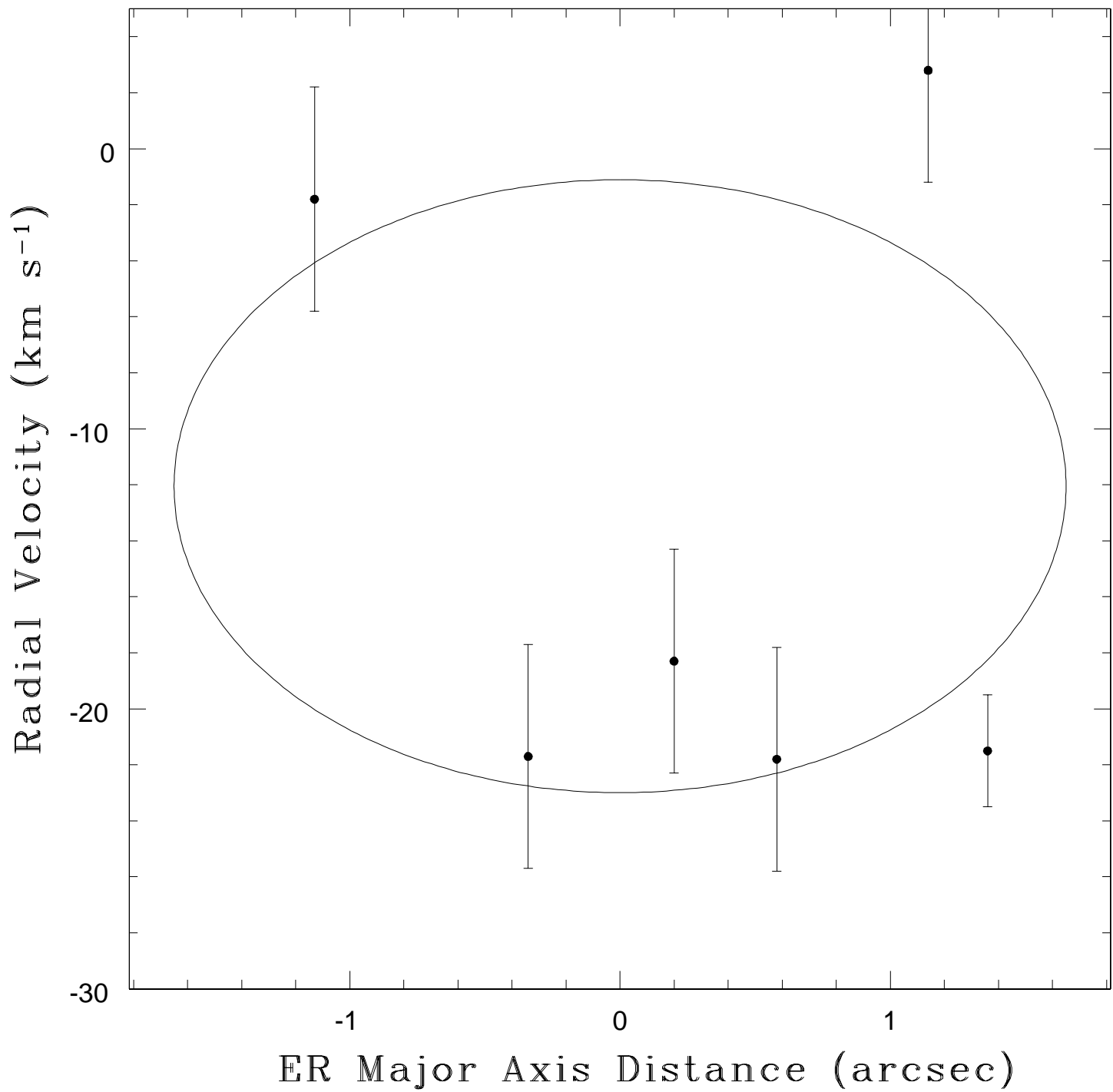
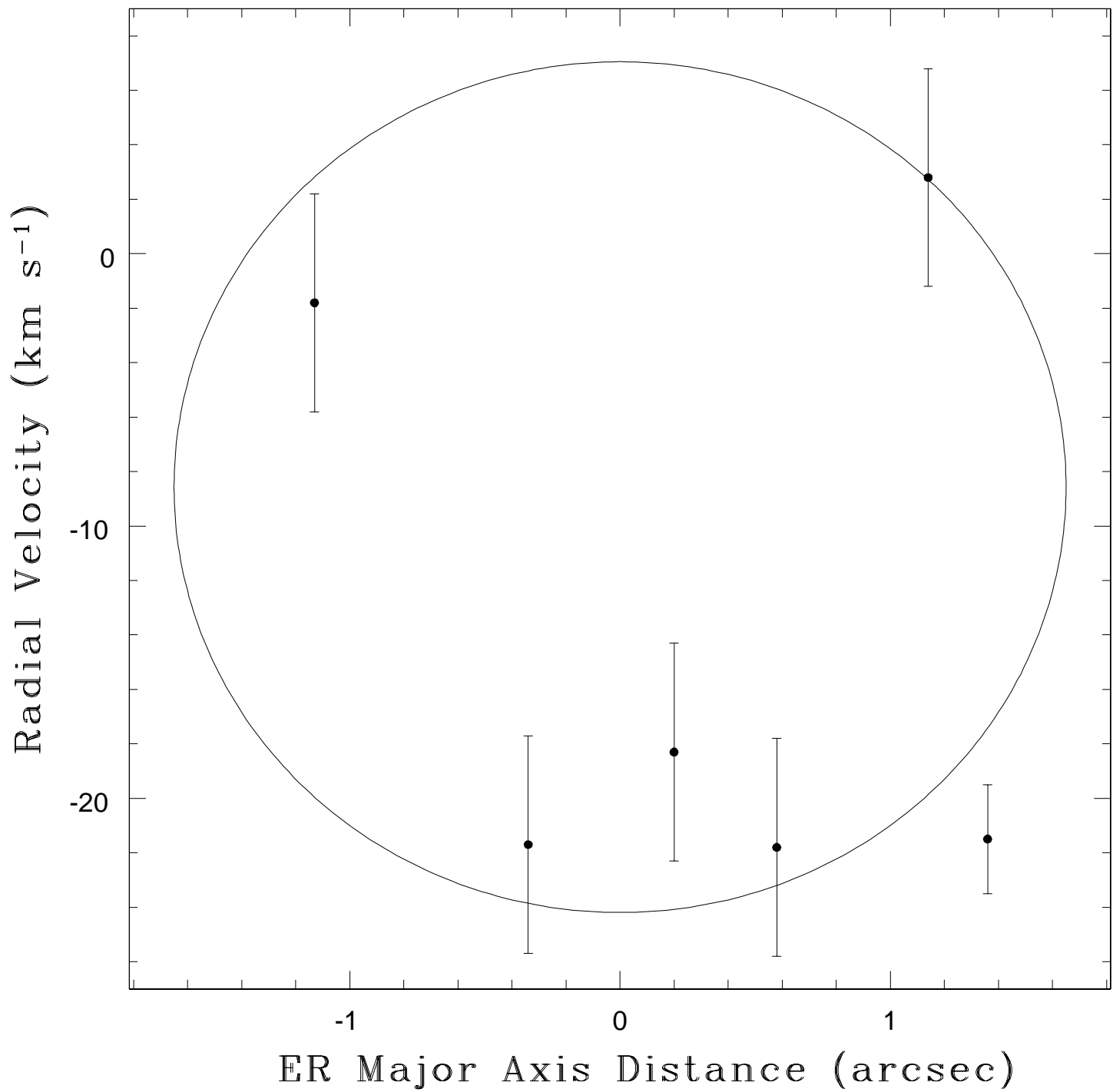
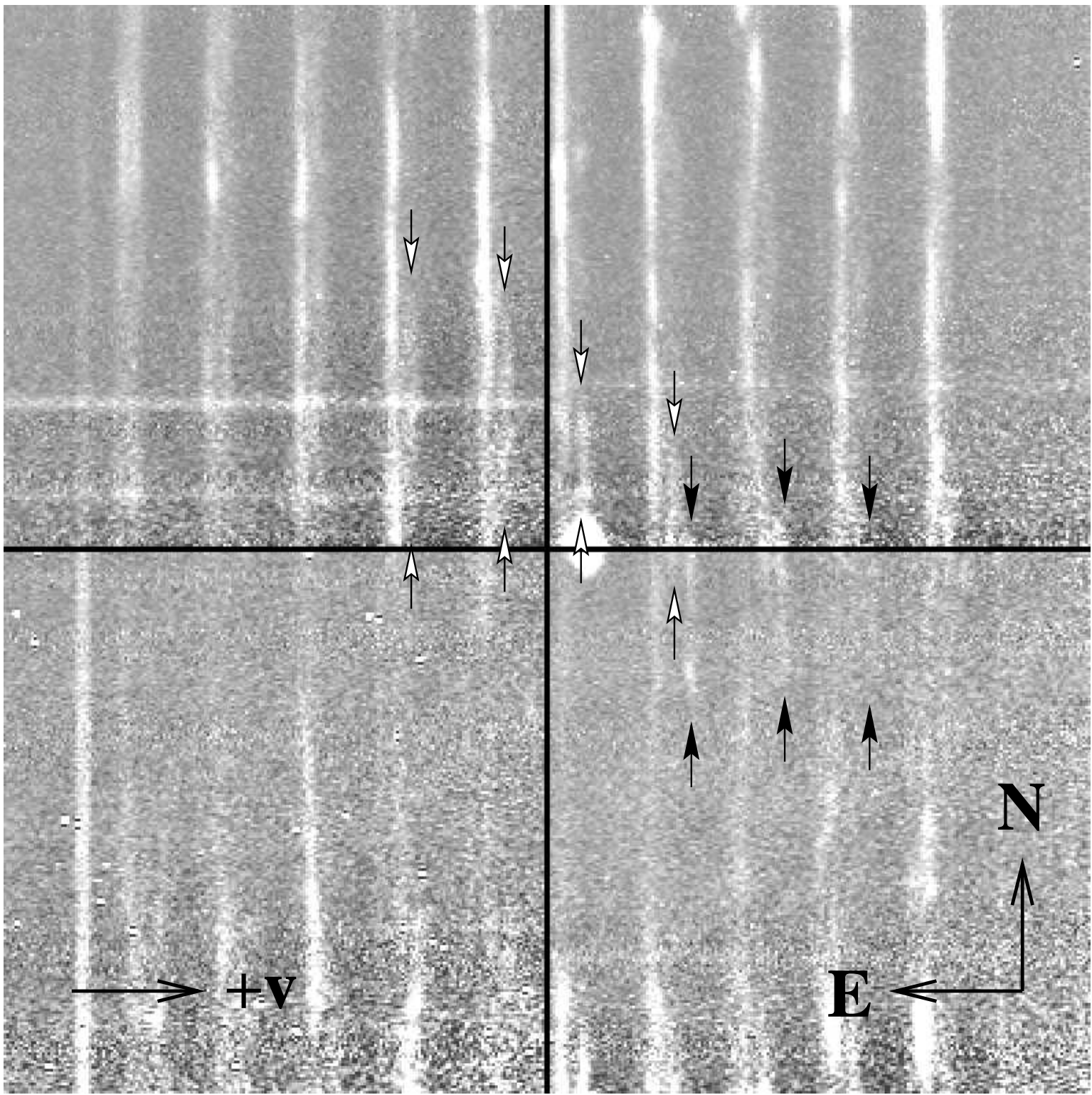
\end{document}